\documentclass[12pt]{article}
\usepackage{amsmath}
\usepackage{amssymb}
\usepackage{graphicx}
\usepackage{caption2}
\usepackage{amsfonts}
\usepackage{cite}

\newcommand{\be}{\begin{equation}}
\newcommand{\ee}{\end{equation}}
\newcommand{\bea}{\begin{eqnarray}}
\newcommand{\eea}{\end{eqnarray}}
\newcommand{\bef}{\begin{figure}}
\newcommand{\ef}{\end{figure}}
\newcommand{\bt}{\begin{tabular}}
\newcommand{\et}{\end{tabular}}
\newcommand{\bno}{\begin{enumerate}}
\newcommand{\eno}{\end{enumerate}}

\def\3{\ss}

\catcode`\"=\active
\def"{\accent'177}
\begin{document}

\begin{center}

{\Large\bf Ultra-chaotic property of Navier-Stokes turbulence}

\vspace{0.3cm}

Shijie Qin$^{1,2}$, Kun Xu$^{1}$ and Shijun Liao$^{2,3 *}$

$^{1}$Dept. of Mathematics, Hong Kong University of Science and Technology, Hong Kong, China\\

$^{2}$State Key Laboratory of Ocean Engineering, Shanghai 200240, China\\

$^{3}$School of Ocean and Civil Engineering, Shanghai Jiao Tong University, Shanghai 200240, China\\

\vspace{0.3cm}

$^*$Corresponding author: sjliao@sjtu.edu.cn

\end{center}

\hspace{-0.5cm}{\bf Abstract} 
{\em A chaotic system is called ultra-chaos when its statistics have sensitivity dependence on initial condition and/or other small disturbances. In this paper, using two-dimensional turbulent Kolmogorov flow as an example, we illustrate that tiny variation of initial condition of Navier-Stokes equations can lead to huge differences not only in spatiotemporal trajectory but also in flow symmetry and its statistics. Here, in order to avoid the influence of artificial numerical noise, we apply ``clean numerical simulation'' (CNS) which can guarantee that the numerical noise can be reduced to such a desired low level that they are negligible in a time interval long enough for calculating statistics. This discovery highly suggests that the Navier-Stokes turbulence (i.e. turbulence governed by the Navier-Stokes equations) might be an ultra-chaos, say, small disturbances must be considered even from viewpoint of statistics. This however leads to a paradox in logic, since small disturbances, which are unavoidable in practice, are unfortunately neglected by the Navier-Stokes turbulence.  Some fundamental characteristics of  turbulence model are discussed and suggested in general meanings.                      
}

%\begin{keyword}
\hspace{-0.5cm}{\bf Keywords} Navier-Stokes turbulence, ultra-chaos, clean numerical simulation
%\end{keyword}

\section{Introduction}

The Navier-Stokes (NS) equations are widely used as a mathematical model to describe turbulent flows.  This model  is so important in physics and engineering that it has been listed as one of the seven Millennium Prize Problem \cite{MillenniumProblem}.  Many researchers, such as Deissler~\cite{Deissler1986PoF}, Bffetta \& Musacchio~\cite{boffetta2017chaos}, and Berera \& Ho~\cite{berera2018chaotic},  reported that spatiotemporal trajectories of the NS equations have sensitivity dependence on initial conditions, corresponding to ``trajectory instability''.  In 2023, Ge, Rolland and Vassilicos~ \cite{Vassilicos2023JFM}   pointed out that the uncertainty of a three-dimensional NS turbulence (i.e. turbulent flow governed by the NS equations) grows exponentially.  All of these provide us rigorous evidences that the NS turbulence is chaotic, say, all tiny (natural/artificial) stochastic disturbances increase exponentially to a macroscopic level of spatiotemporal trajectories due to the famous ``butterfly-effect'' of chaos  \cite{Lorenz1963}.  

It is well known that, for turbulent flows, statistical results are more important than spatio-temporal trajectories. Normally, although the trajectory of a chaotic system is sensitive to small disturbance, its statistic results are stable. Such kind of chaos is called ``normal-chaos''. However, statistical results of some chaotic systems are even sensitive to small disturbances, corresponding to ``statistic instability'', called ``ultra-chaos'' \cite{Liao2022AAMM, Yang2023CSF, qin_liao_2023, Zhang2023PhysicaD}. Obviously, ultra-chaos has higher disorder than normal-chaos.  

In this paper, using the two-dimensional (2D) turbulent Kolmogorov flow as an example, we illustrate in \S~3 that the Navier-Stokes turbulence might be ultra-chaos, say, even its statistic results might have sensitivity dependence on initial condition. It means that small disturbances {\em cannot} be neglected for the Navier-Stokes turbulence even from viewpoint of {\em statistics}.  But unfortunately, tiny stochastic disturbances are in fact indeed neglected in the Navier-Stokes equations.   This certainly leads to a serious paradox in logic.  Some fundamental characteristics of a turbulence model in general are discussed in \S~4.             

\section{Mathematical model of Navier-Stokes turbulence}
Consider the 2D incompressible Kolmogorov flow \cite{obukhov1983kolmogorov, chandler2013invariant, wu2021quadratic} in a square domain $[0, L]^2$ with the periodic boundary condition in space, under the so-called Kolmogorov forcing that is stationary, monochromatic, and periodically varying in space with an integer $n_K$ describing the forcing scale and $\chi$  the  forcing amplitude per unit mass of fluid, respectively.  Using the length scale $L/2\pi$ and the time scale $\sqrt{L/2\pi\chi}$, the governing equation of this 2D Kolmogorov flow, say, the non-dimensional Navier-Stokes equation in the form of stream function,  reads 
\begin{equation}
 \frac{\partial}{\partial t}\Big(\nabla^{2}\psi\Big)+\frac{\partial(\psi,\nabla^{2}\psi)}{\partial(x,y)}-\frac{1}{Re}\nabla^{4}\psi+n_K\cos(n_Ky)=0,       \label{eq_psi}
\end{equation}
subject to the periodic boundary condition
\begin{equation}
\psi(x, y, t)=\psi(x+2\pi, y, t)=\psi(x, y+2\pi, t),       \label{boundary_condition}
\end{equation}
where 
\[
Re=\frac{\sqrt{\chi}}{\nu}\left(\frac{L}{2\pi}\right)^{\frac{3}{2}}   
\]
is the Reynolds number, $\nu$ denotes the kinematic viscosity, $\psi$ is the stream function defined by 
\[
u=-\frac{\partial\psi}{\partial y}, \hspace{0.75cm} v=\frac{\partial\psi}{\partial x},
\]
$u$ and $v$ represent horizontal and vertical velocities, $t$ denotes the time,  $x,y\in[0,2\pi]$ are horizontal and vertical coordinates, $\nabla^{2}$ is the Laplace operator,  $\nabla^{4}=\nabla^{2}\nabla^{2}$, and 
\[
 \frac{\partial(a,b)}{\partial(x,y)}=\frac{\partial a}{\partial x}\frac{\partial b}{\partial y}-\frac{\partial b}{\partial x}\frac{\partial a}{\partial y}      
\]
is the Jacobi operator, respectively.  In order to investigate a relatively strong state of turbulent flow, we choose $n_K=16$ and $Re=2000$ for all cases considered in this paper.  

To investigate its sensitivity dependence on initial condition, let us consider the following three different initial conditions:
\begin{eqnarray}
 \Phi_{1,0}(x,y)=\psi(x,y,0) & = & -\frac{1}{2}\big[\cos(4x+4y) + \cos(4x-4y) + \sin(4x+4y)\big] \nonumber\\&+& 10^{-10}\big[\sin(x+y) + \sin(x-y)\big],       \label{initial_condition-1} \\
 \Phi_{2,0}(x,y)= \psi(x,y,0) & = & -\frac{1}{2}\big[\cos(4x+4y) + \cos(4x-4y) + \sin(4x+4y)\big] \nonumber\\&+& 10^{-10}\big[\sin(2x+y) + \sin(2x-y)\big],       \label{initial_condition-2} \\
  \Phi_{3,0}(x,y)= \psi(x,y,0) & = & -\frac{1}{2}\big[\cos(4x+4y) + \cos(4x-4y) + \sin(4x+4y)\big] \nonumber\\&+& 10^{-10}\big[\sin(4x+y) + \sin(4x-y)\big],       \label{initial_condition-3}
\end{eqnarray}
corresponding to three turbulent Kolmogorov flows with different spatial symmetries, as described below.   Note that the deviations  between these three initial conditions  $\Phi_{1,0}(x,y)$, $\Phi_{2,0}(x,y)$, and $\Phi_{3,0}(x,y)$ are quite small over the whole field, i.e. 
\begin{equation} 
\left|\Phi_{m,0}(x,y)-\Phi_{n,0}(x,y) \right| \leq 4 \times 10^{-10}, \hspace{0.75cm} x,y \in [0,2 \pi], \label{small-difference-IC}
\end{equation}
where $m, n = 1,2,3$ but $m\neq n$. The spatiotemporal evolutions of such kind of tiny deviations are neglected by some traditional numerical algorithms (even including the direct numerical simulation, i.e. DNS), but can be accurately obtained by clean numerical simulation (CNS) \cite{Liao2009, Liao2023book, Hu2020JCP, Qin2020CSF, Qin2022JFM,  Liao-2025-JFM-NEC, Liao-2025-JFM-PS}, which will be briefly described in \S~3.      

It should be emphasized that the initial condition $ \Phi_{1,0}(x,y)$ given by (\ref{initial_condition-1}) has the spatial translation symmetry~A:   
\begin{equation}
{S}(x,y) = S(x+\pi,y+\pi),    \label{symmetry-1}
\end{equation}
 the initial condition $ \Phi_{2,0}(x,y)$ given by (\ref{initial_condition-2}) has the spatial translation symmetry~B: 
\begin{equation}
S(x,y) =S(x+\pi/2,y+\pi),   \label{symmetry-2}
\end{equation}
and  the initial condition $ \Phi_{3,0}(x,y)$ given by (\ref{initial_condition-3})  has the spatial translation symmetry~C:
\begin{equation}
S(x, y)=S(x+\pi/2, y),   \label{symmetry-3}
\end{equation}
respectively.  It should be emphasized that the above-mentioned  three spatial symmetries are quite different.  Note that the vorticity $\omega=\nabla^{2}\psi$  retains the same spatial symmetry as the stream function $\psi$.
 
\section{Ultra-chaotic property of Navier-Stokes turbulence}

As pointed out by many researchers \cite{Deissler1986PoF, boffetta2017chaos, berera2018chaotic, Vassilicos2023JFM}, Navier-Stokes turbulence (i.e. turbulent flows governed by NS equations) are chaotic.  Note that numerical noises are avoidable for all algorithms including DNS.  So, due to the butterfly-effect of chaos \cite{Lorenz1963}, spatiotemporal trajectories given by traditional numerical  algorithms including DNS are quickly polluted by numerical noises, i.e. far away from the true solution of NS turbulence. To obtian an accurate enough spatiotemporal trajectory of the NS turbulence, whose numerical noise is much smaller than that of the true solution  and thus is negligible in a long enough interval of time, we use the CNS \cite{Liao2009, Liao2023book, Hu2020JCP, Qin2020CSF, Qin2022JFM,  Liao-2025-JFM-NEC, Liao-2025-JFM-PS} to solve Eq.~(\ref{eq_psi}) subject to the periodic boundary condition (\ref{boundary_condition}) and one initial condition among (\ref{initial_condition-1} )-(\ref{initial_condition-3}).  The three CNS results are named by Flow~CNS-1 corresponding to  (\ref{initial_condition-1}), Flow~CNS-2  corresponding to  (\ref{initial_condition-2}), and Flow~CNS-3  corresponding to  (\ref{initial_condition-3}), respectively.             

For the sake of simplicity, the basic ideas of the CNS \cite{Liao2009, Liao2023book, Hu2020JCP, Qin2020CSF, Qin2022JFM, Qin2024JOES, Liao-2025-JFM-NEC, Liao-2025-JFM-PS} are briefly described here. Firstly, to decrease the spatial truncation error to a small enough level, we use an uniform mesh $N^2 = 1024^2$ and adopt the Fourier pseudo-spectral method for spatial approximation with the $3/2$ rule for dealiasing. In this way, the corresponding spatial resolution is fine enough for the considered three kinds of turbulent Kolmogorov flows: the grid spacing is less than the averaged Kolmogorov scale and enstrophy dissipative scale, according to  Pope~\cite{pope2001turbulent} and Boffetta \& Ecke~\cite{Boffetta2012ARFM}, which is carefully checked and confirmed in this paper.  
But unlike DNS, in order to decrease the temporal truncation error to a small enough level, we use the $M$th-order Taylor expansion with the time step $\Delta t = 10^{-3}$. Especially, we use {\em multiple-precision} with $N_s$ significant digits for all physical/numerical variables and parameters so as to decrease the round-off error to a small enough level. In addition, another CNS result is given by the same CNS algorithm but with even smaller numerical noise, i.e. using even larger $M$ and $N_{s}$ than those mentioned above, which guarantees (by comparison) that the numerical noise of the former CNS result is rigorously negligible throughout the whole time interval $t\in[0,300]$ so that it can be used as a ``clean'' benchmark solution of the NS turbulence. It is found that we need to adopt $M=60$ \& $N_{s}=110$ for Flow~CNS-1, $M=100$ \& $N_{s}=230$ for Flow~CNS-2, and $M=120$ \& $N_{s}=430$ for Flow~CNS-3 in the same time interval $t\in[0,300]$, respectively, because the three different turbulent flows  have different noise-growing exponents so that different values of $M$ and $N_{s}$ in the frame of CNS are necessary to decrease their respective numerical noises to the corresponding small enough levels. Note that the self-adaptive strategy \cite{Qin2023AAMM} and parallel computing are adopted to dramatically increase the computational efficiency of the CNS algorithm.  For further details about the CNS algorithm mentioned above, please refer to Qin et al.~\cite{Qin2024JOES} and the Liao's book \cite{Liao2023book}.  %Note that the related code of CNS can be downloaded via Github.
The detailed comparisons of spatial symmetry and statistical results between the three different turbulent flows, i.e. Flow~CNS-1,  Flow~CNS-2, and Flow~CNS-3 are described below. 
 
\begin{figure}
    \begin{center}f
        \begin{tabular}{cc}
             \includegraphics[width=2.0in]{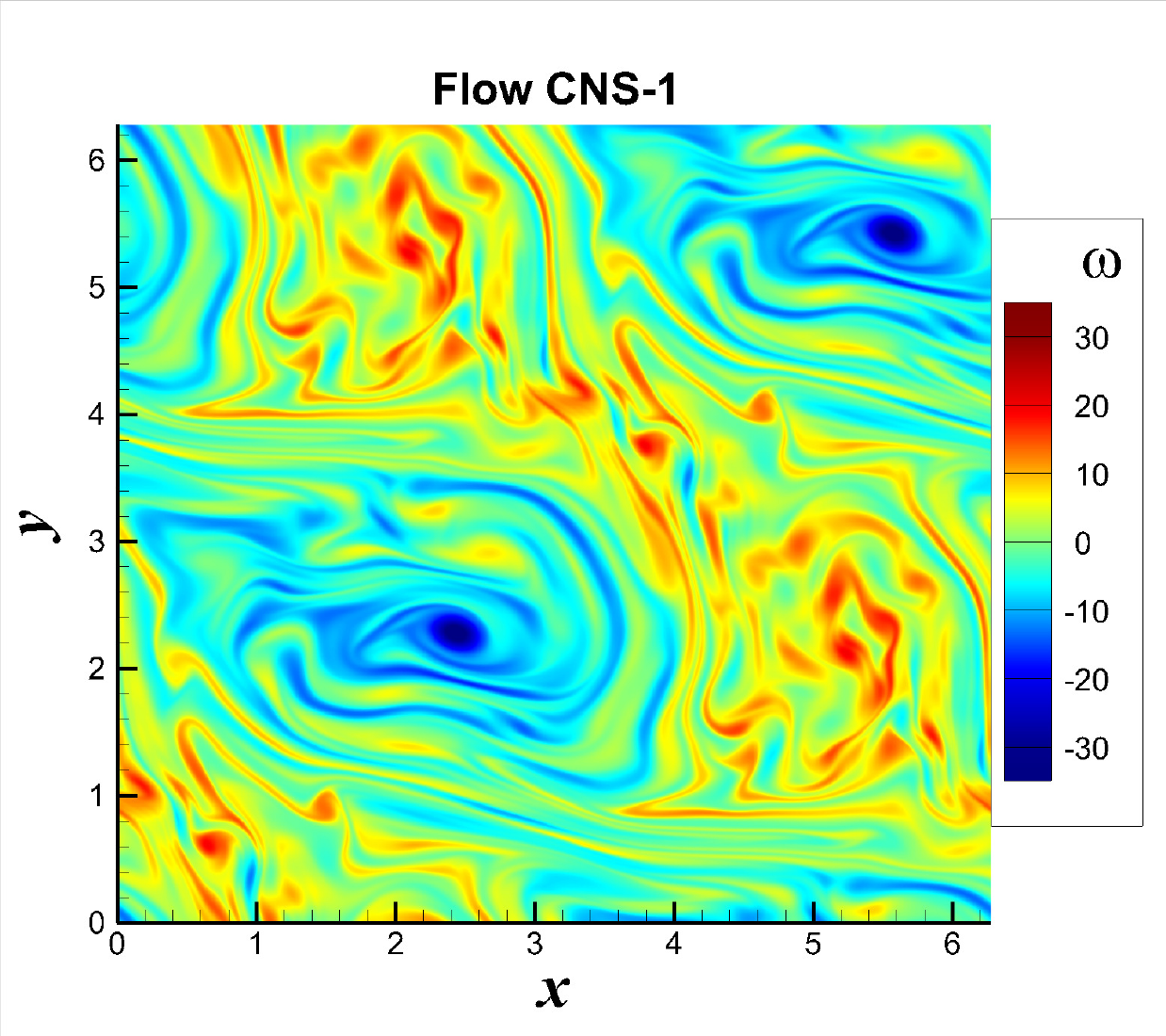}
             \includegraphics[width=2.0in]{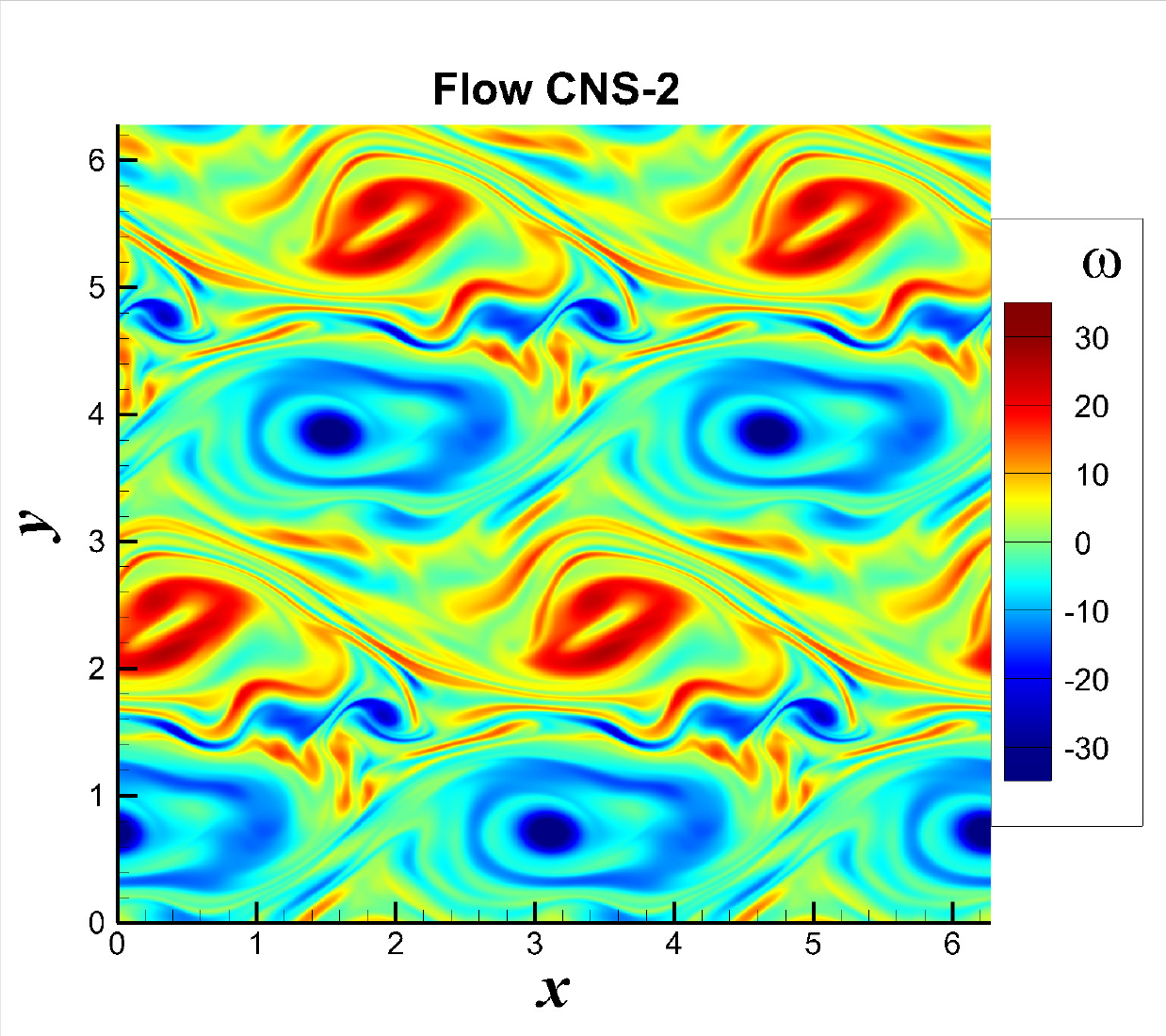}
             \includegraphics[width=2.0in]{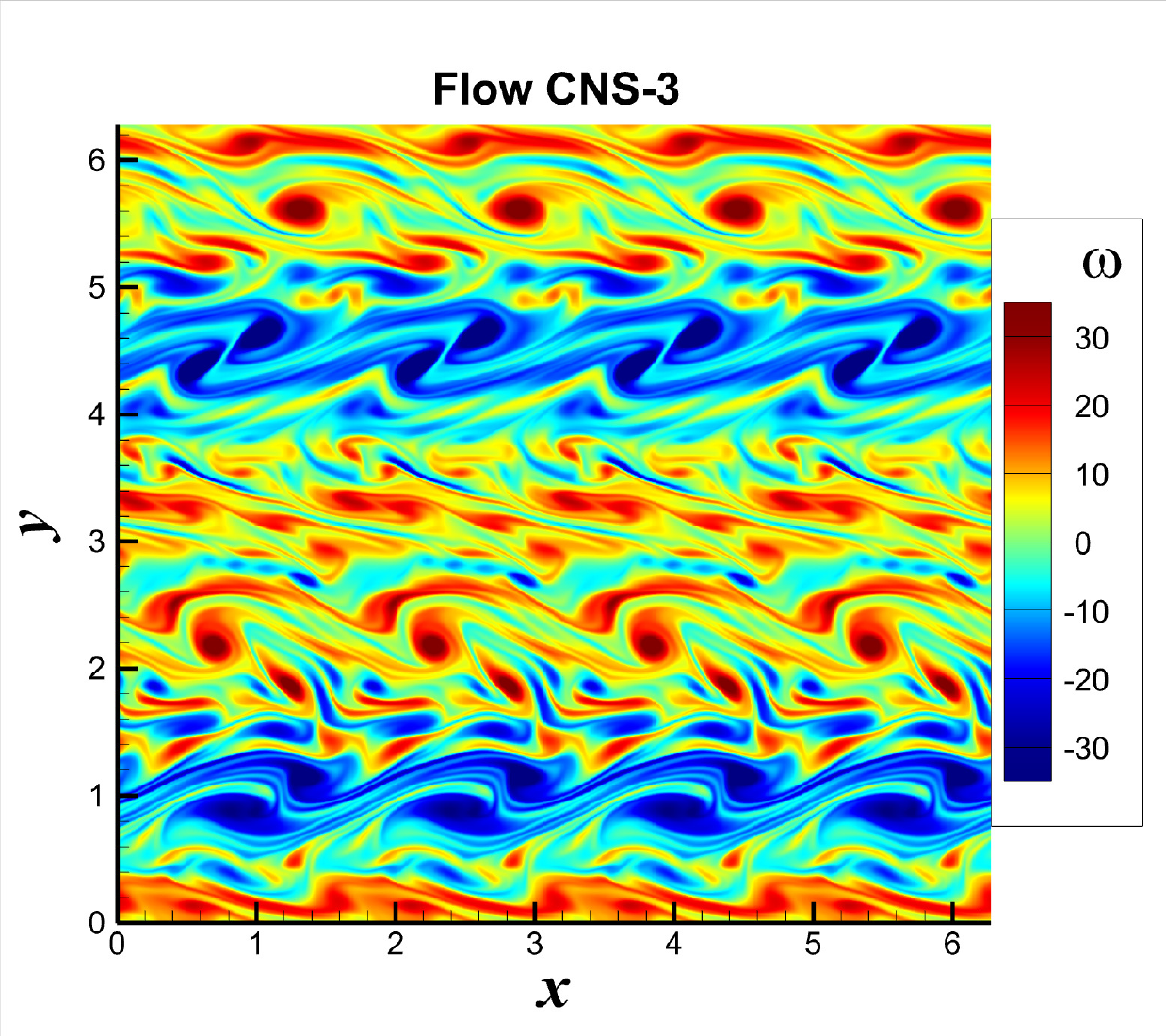}
        \end{tabular}
    \caption{Vorticity fields $\omega$ at $t=100$ of the 2D turbulent Kolmogorov flow governed by (\ref{eq_psi}) and (\ref{boundary_condition}) in the case of $n_K=16$ and $Re=2000$, given by CNS subject to the initial conditions (\ref{initial_condition-1}) (left, marked by Flow~CNS-1), (\ref{initial_condition-2}) (middle, marked by Flow~CNS-2), and (\ref{initial_condition-3}) (right, marked by Flow~CNS-3), respectively.
}     \label{Contour}
    \end{center}
\end{figure}

Firstly, let us  compare  the  spatial  symmetry of the flow field.   It is found that the vorticity field $\omega$ of Flow~CNS-1  remains the {\em same} spatial translation symmetry (\ref{symmetry-1}) as its initial condition (\ref{initial_condition-1}) throughout the {\em whole} time interval $t\in[0,300]$, for example at $t=100$ as shown in figure~\ref{Contour}(a). Similarly, the vorticity field $\omega$ of Flow~CNS-2 remains the same spatial translation symmetry (\ref{symmetry-2}) as the initial condition (\ref{initial_condition-2}) throughout the whole time interval $t\in[0,300]$, and the vorticity field $\omega$ of Flow~CNS-3 remains the same spatial translation symmetry (\ref{symmetry-3}) as the initial condition (\ref{initial_condition-3}) throughout the whole time interval $t\in[0,300]$, for examples at $t=100$ as shown in figures~\ref{Contour}(b) and (c), respectively. Here, we had to emphasize that, as shown by (\ref{small-difference-IC}), the deviations between the three initial conditions (\ref{initial_condition-1})-(\ref{initial_condition-3}) are quite small, which however lead to three completely distinct turbulent flows with different spatial symmetries. This clearly illustrates that the spatial symmetry of the Navier-Stokes turbulence has sensitivity dependence on the initial condition.

\begin{figure}
    \begin{center}
        \begin{tabular}{cc}
             \includegraphics[width=2.0in]{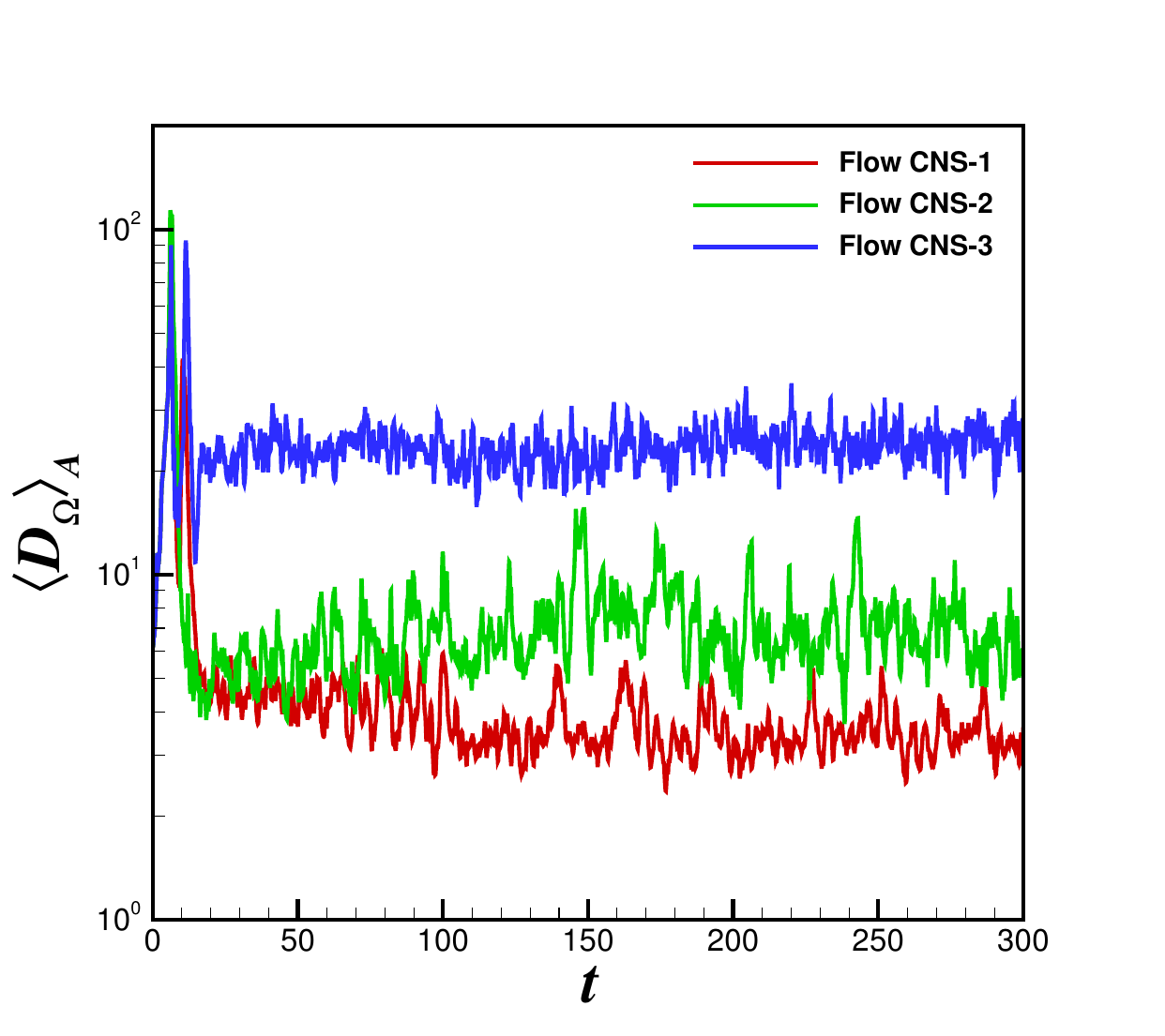}
        \end{tabular}
    \caption{Comparison of time histories of the spatially averaged enstrophy dissipation rate $\langle D_\Omega\rangle_A$ of the 2D turbulent Kolmogorov flow governed by (\ref{eq_psi}) and (\ref{boundary_condition}) in the case of $n_K=16$ and $Re=2000$, given by Flow~CNS-1 (red line), Flow~CNS-2 (green line), Flow~CNS-3 (blue line), respectively.}     \label{Do_t}
    \end{center}
\end{figure}

\begin{figure}
    \begin{center}
        \begin{tabular}{cc}
             \includegraphics[width=2.0in]{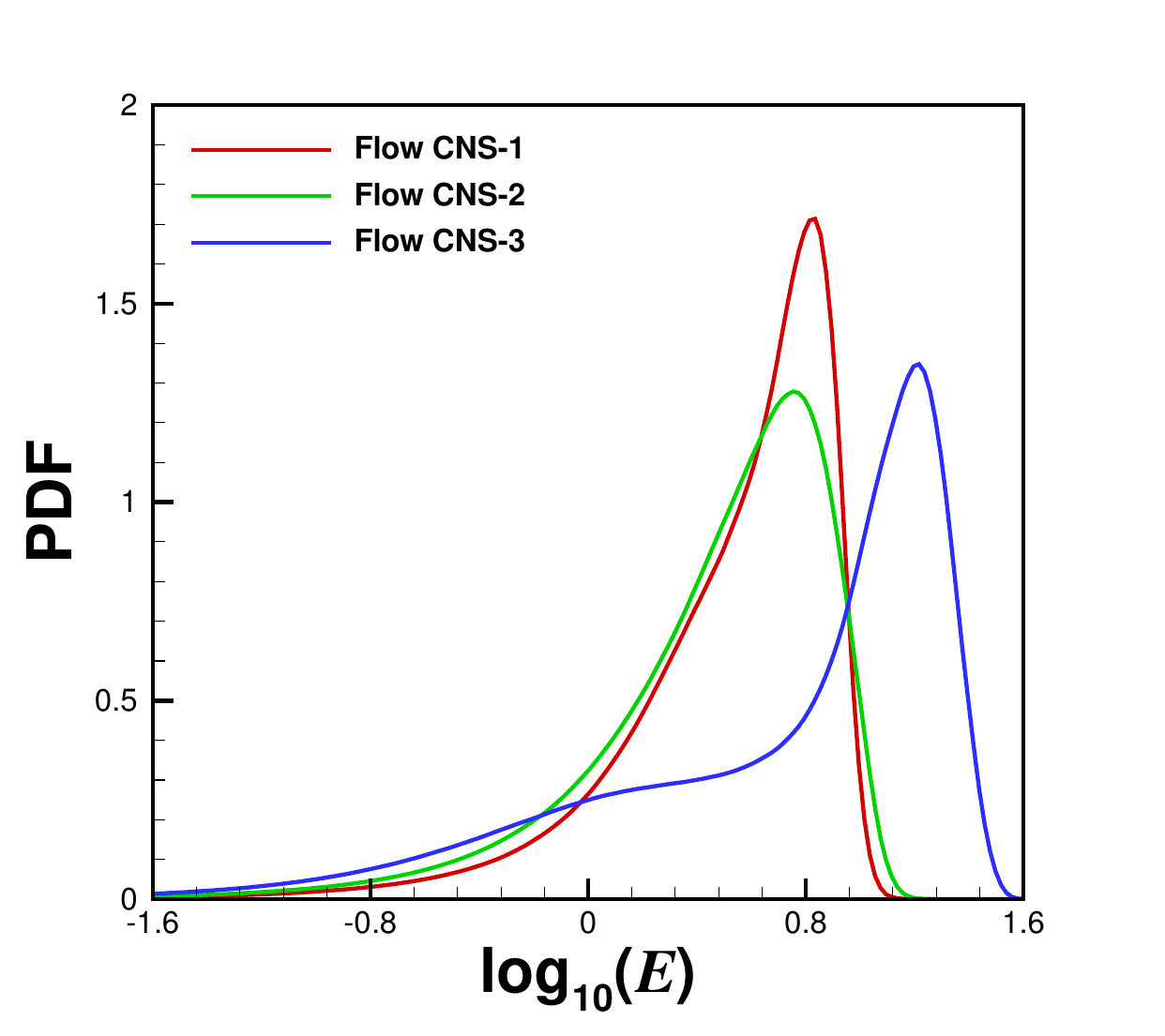}
             \includegraphics[width=2.0in]{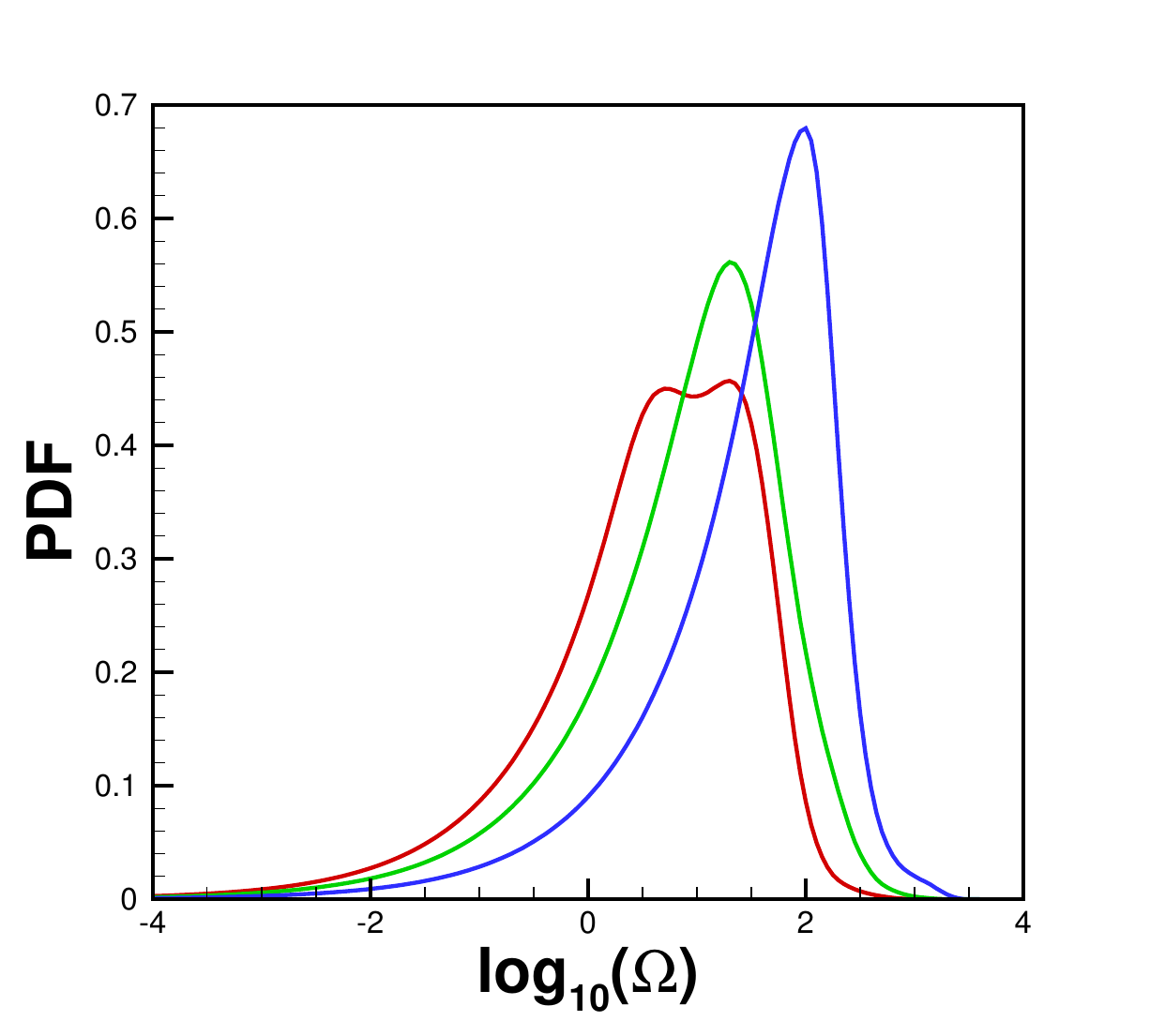}\\
             \includegraphics[width=2.0in]{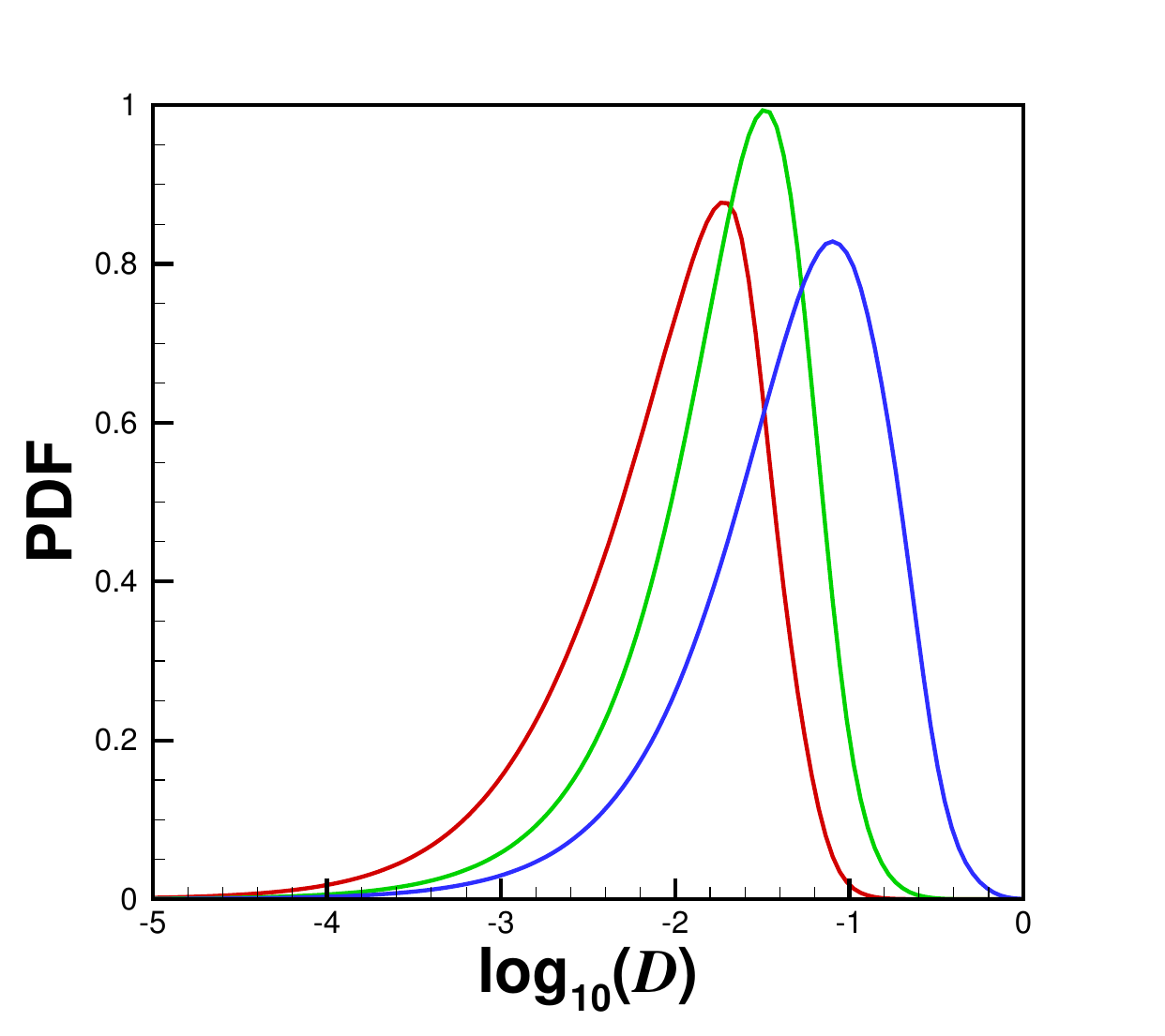}
             \includegraphics[width=2.0in]{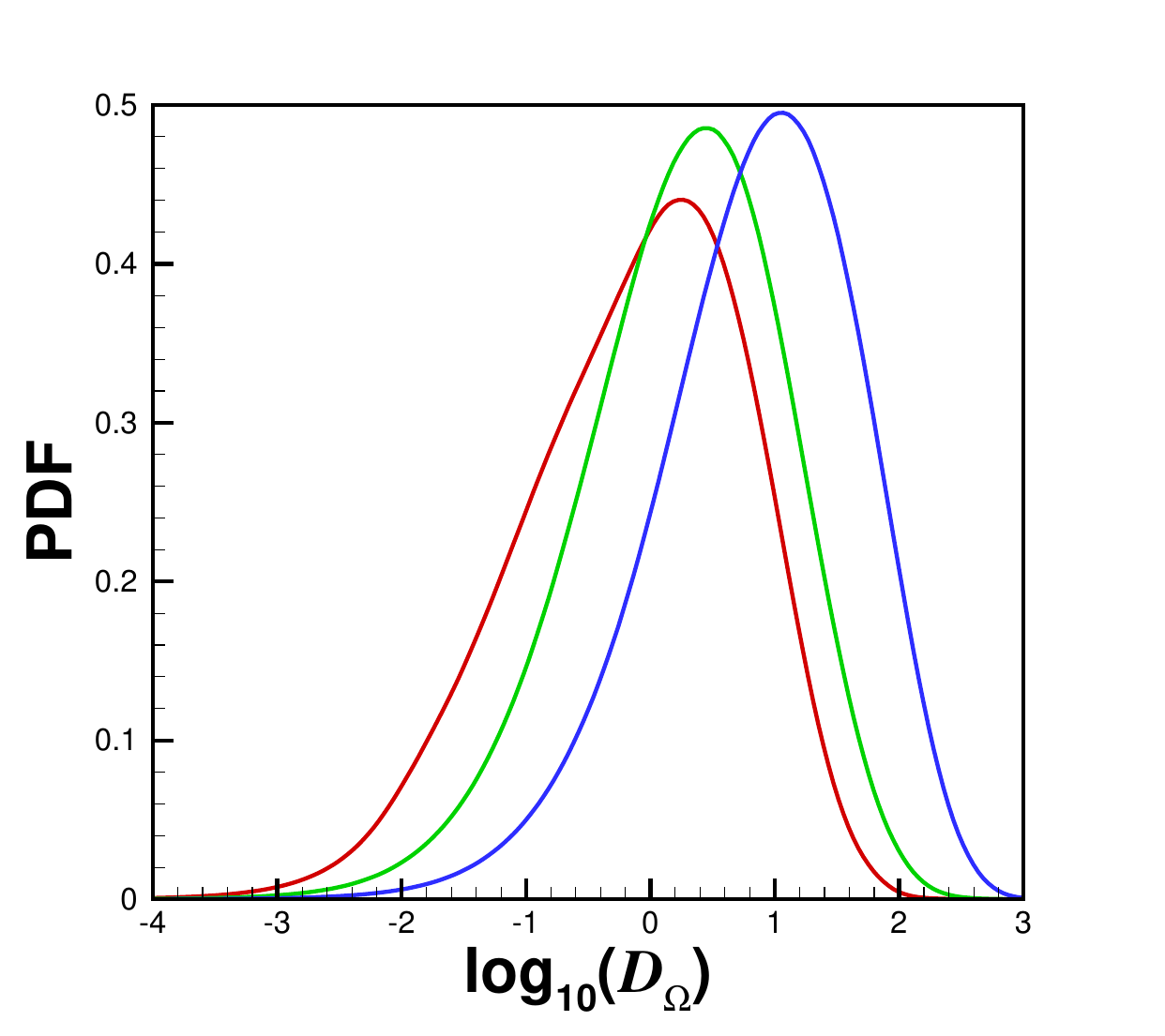}
        \end{tabular}
    \caption{Comparison of probability density functions (PDFs) of (top-left) kinetic energy $E(x,y,t)$, (top-right) enstrophy $\Omega(x,y,t)$, (bottom-left) kinetic energy dissipation $D(x,y,t)$, and (bottom-right) enstrophy dissipation rate $D_\Omega(x,y,t)$ of the 2D turbulent Kolmogorov flow, governed by (\ref{eq_psi}) and (\ref{boundary_condition}) in the case of $n_K=16$ and $Re=2000$, given by Flow~CNS-1 (red line), Flow~CNS-2 (green line), Flow~CNS-3 (blue line), respectively. Their definitions are given in Appendix.}     \label{E-PDF}
    \end{center}
\end{figure}

\begin{figure}
    \begin{center}
        \begin{tabular}{cc}
             \includegraphics[width=2.0in]{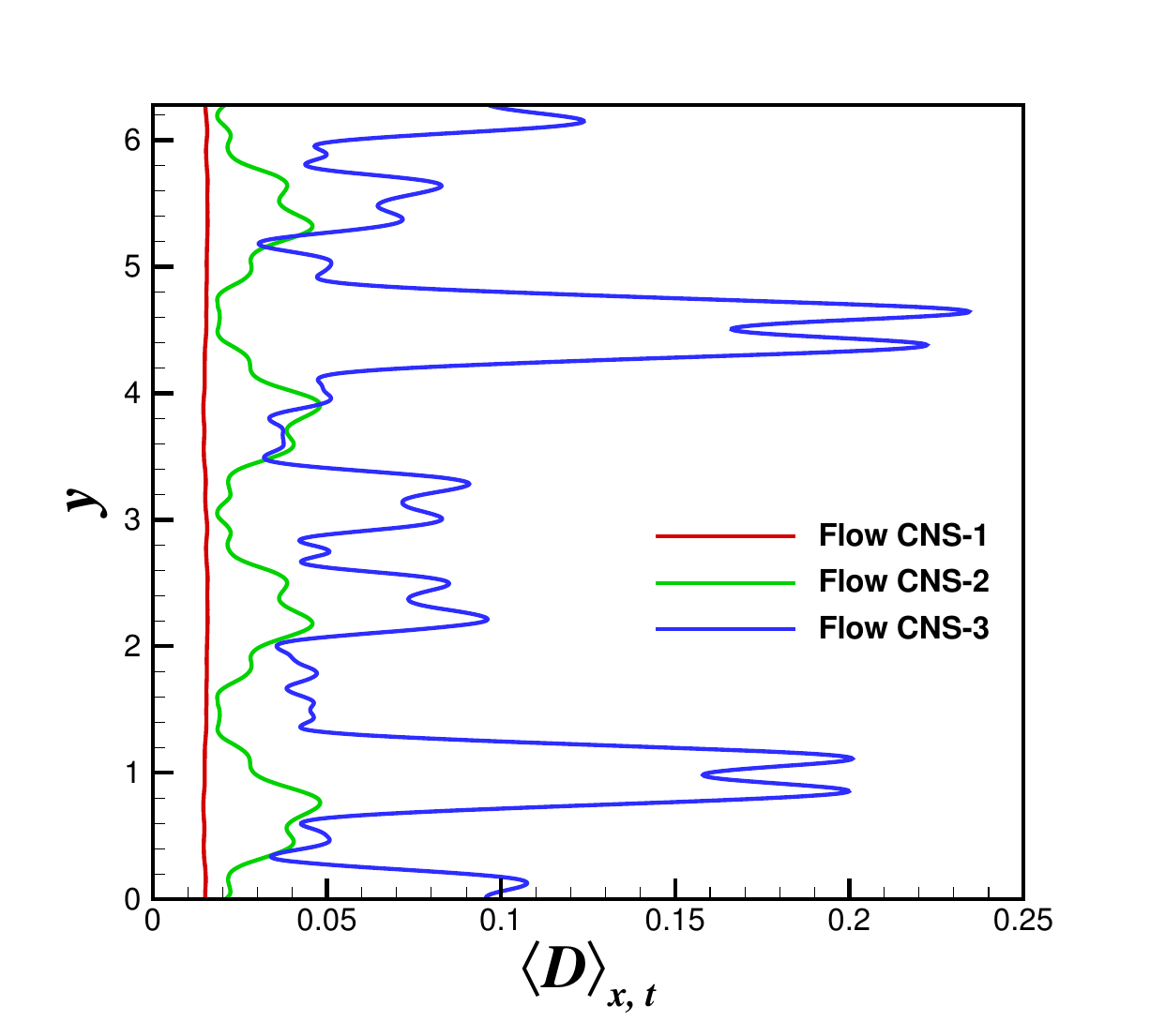}
             \includegraphics[width=2.0in]{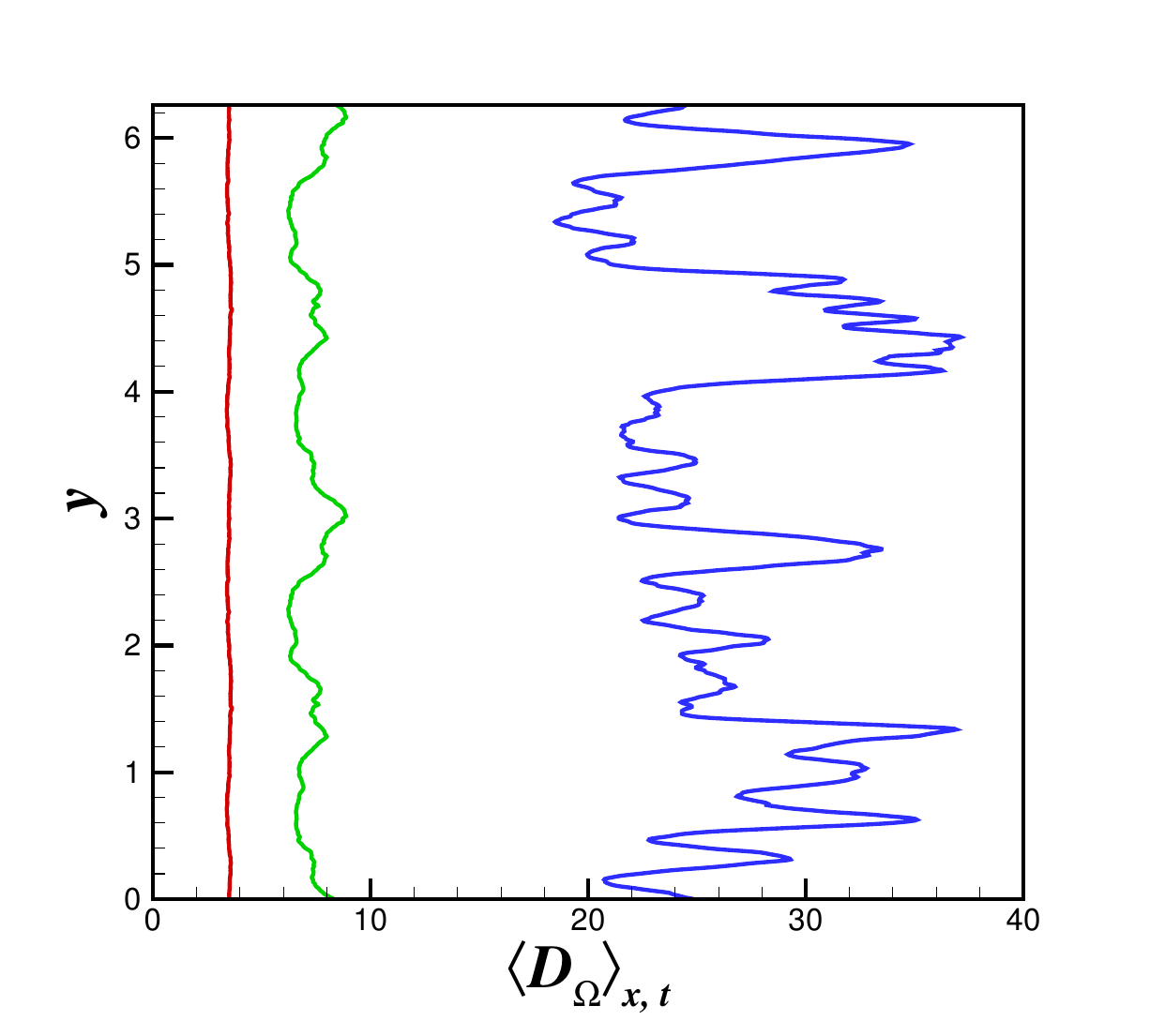}
        \end{tabular}
    \caption{Comparison of the spatiotemporally averaged (left) kinetic energy dissipation rate $\langle D\rangle_{x,t}(y)$ and (right) enstrophy dissipation rate $\langle D_\Omega\rangle_{x,t}(y)$ of the 2D turbulent Kolmogorov flow governed by Eqs.~(\ref{eq_psi}) and (\ref{boundary_condition}) in the case of $n_K=16$ and $Re=2000$, given by Flow~CNS-1 (red line), Flow~CNS-2 (green line), Flow~CNS-3 (blue line), respectively.}     \label{D_y}
    \end{center}
\end{figure}

\begin{figure}
    \begin{center}
        \begin{tabular}{cc}
             \includegraphics[width=2.0in]{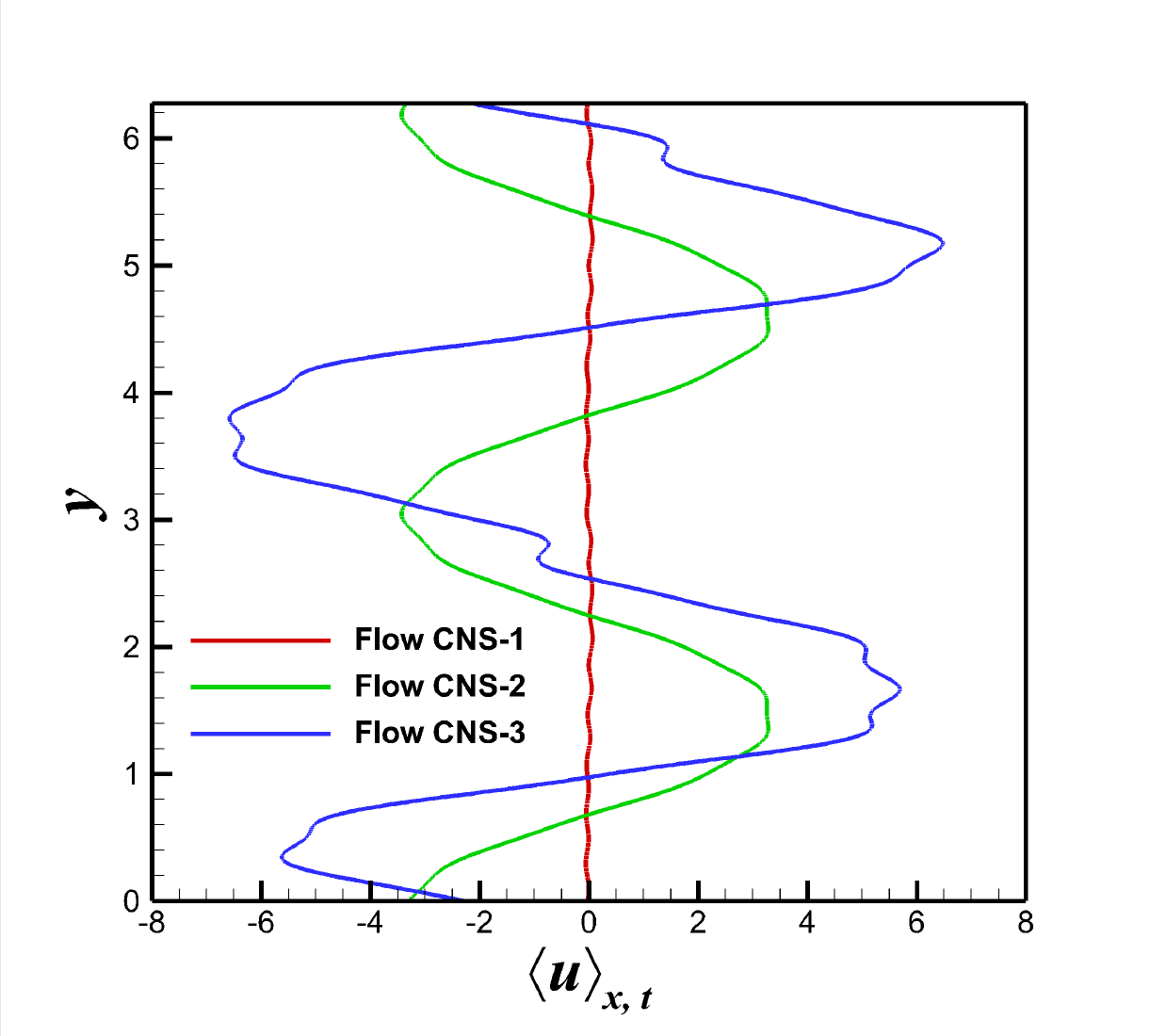}
             \includegraphics[width=2.0in]{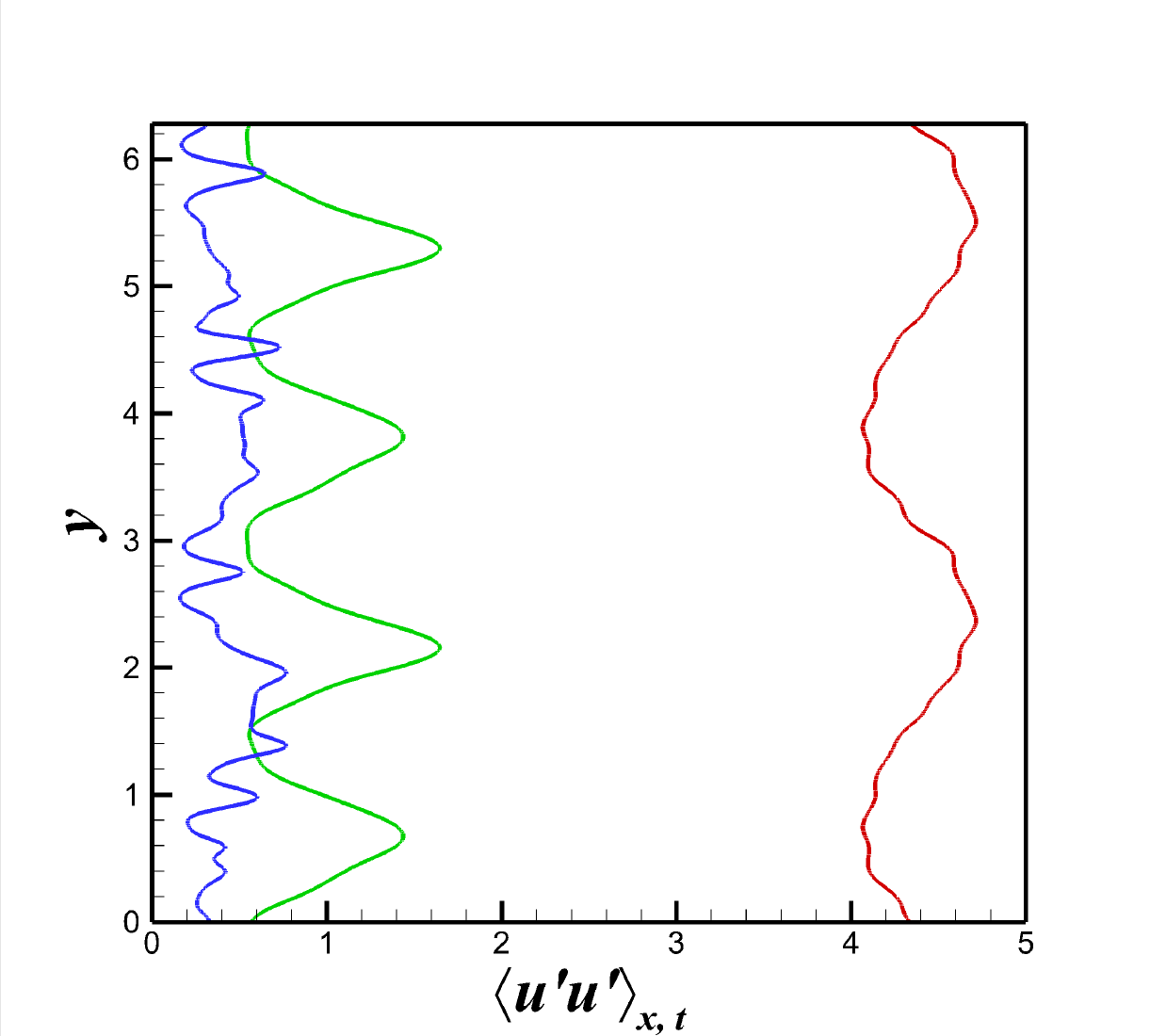}\\
             \includegraphics[width=2.0in]{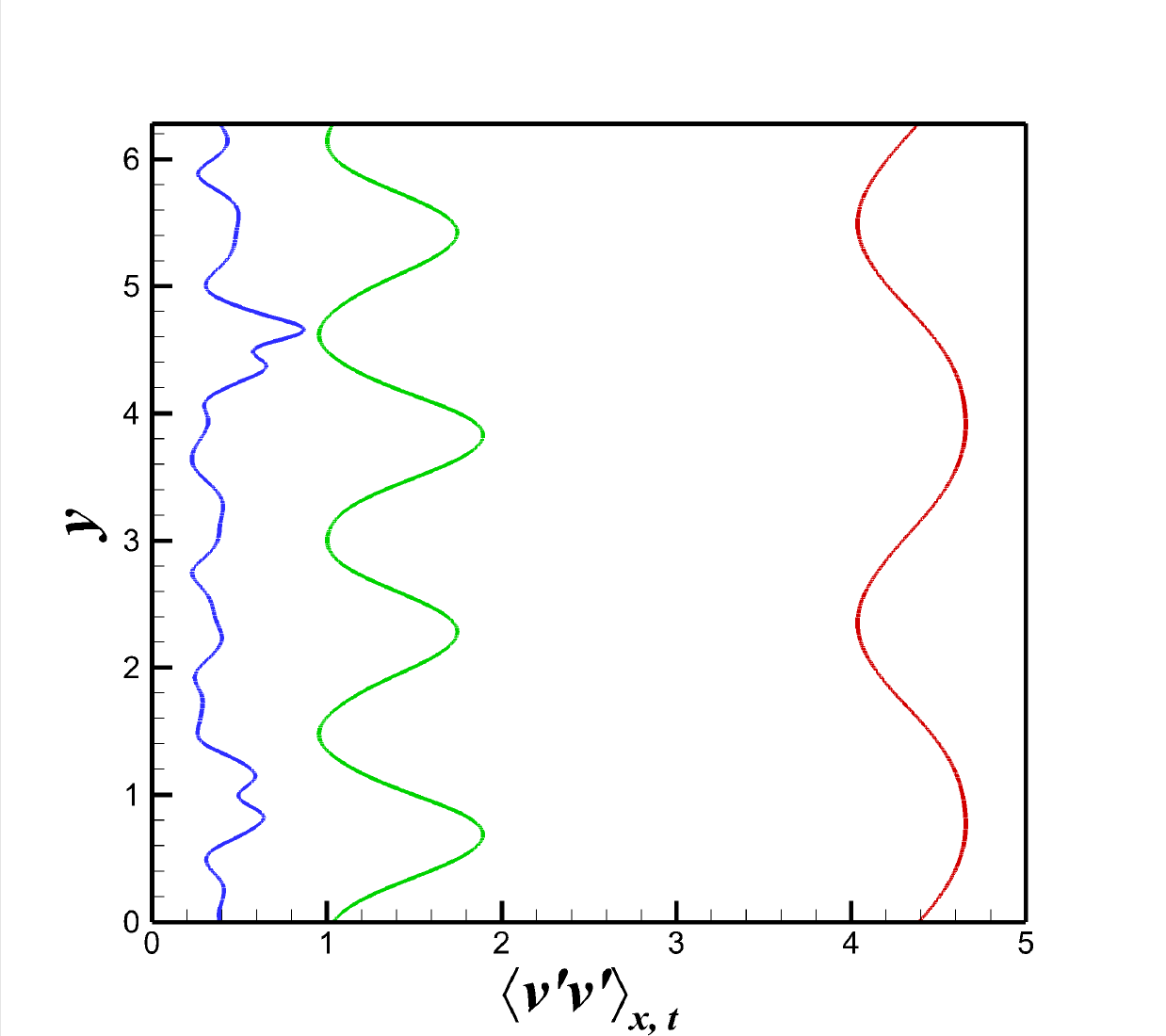}
             \includegraphics[width=2.0in]{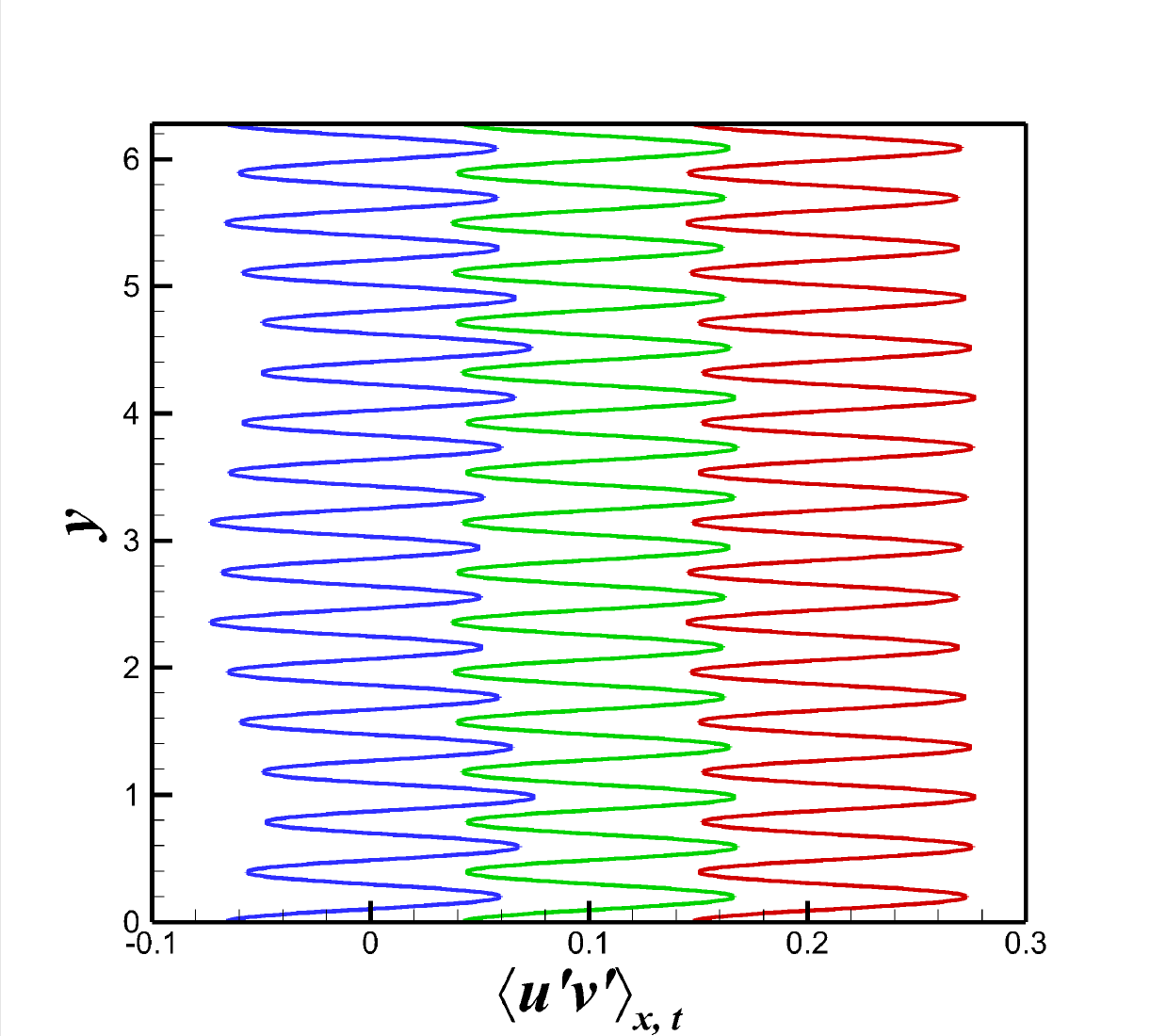}
        \end{tabular}
    \caption{Comparison of the spatiotemporally averaged (top-left) horizontal velocity $\langle u\rangle_{x,t}(y)$, (top-right) normal stresses $\langle u'u'\rangle_{x,t}(y)$,  (bottom-left) normal stresses$\langle v'v'\rangle_{x,t}(y)$, and (bottom-right) shear stress $\langle u'v'\rangle_{x,t}(y)$ of the 2D turbulent Kolmogorov flow, governed by (\ref{eq_psi}) and (\ref{boundary_condition}) in the case of $n_K=16$ and $Re=2000$, given by Flow~CNS-1 (red line), Flow~CNS-2 (green line), Flow~CNS-3 (blue line), respectively. Their definitions are given in Appendix.}     \label{uu_y}
    \end{center}
\end{figure}

Secondly, let us compare some statistic results. Figure~\ref{Do_t} compares the time histories of the spatially averaged enstrophy dissipation rate $\langle D_\Omega\rangle_A$ given by Flow~CNS-1, Flow~CNS-2, and Flow~CNS-3. It is found that the obvious deviations between these time histories appear at $t\approx10$ when the tiny disturbances caused by $10^{-10}[\sin(x+y) + \sin(x-y)]$ in the initial condition (\ref{initial_condition-1}), $10^{-10}[\sin(2x+y) + \sin(2x-y)]$ in the initial condition (\ref{initial_condition-2}), and $10^{-10}[\sin(4x+y) + \sin(4x-y)]$ in the initial condition (\ref{initial_condition-3}) just increase to the macro level.
Because of the different spatial symmetries of the three turbulent flows, there are obvious deviations between the probability density functions (PDFs) of the kinetic energy $E(x,y,t)$, enstrophy $\Omega(x,y,t)$, kinetic energy dissipation $D(x,y,t)$, and enstrophy dissipation rate $D_\Omega(x,y,t)$, as illustrated in figures~\ref{E-PDF}(a) to (d), respectively.
Besides, as shown in figures~\ref{D_y}(a) and (b), the spatiotemporally averaged kinetic energy dissipation rate $\langle D\rangle_{x,t}(y)$ and enstrophy dissipation rate $\langle D_\Omega\rangle_{x,t}(y)$ of Flow~CNS-1, Flow~CNS-2, and Flow~CNS-3 are also quite different. 
Moreover, as illustrated in figures~\ref{uu_y}(a) to (d), there exist the significant deviations between the results of Flow~CNS-1, Flow~CNS-2, and Flow~CNS-3 for the spatiotemporally averaged horizontal velocity $\langle u\rangle_{x,t}(y)$ and Reynolds-stress associated terms $\langle u'u'\rangle_{x,t}(y)$, $\langle v'v'\rangle_{x,t}(y)$, and $\langle u'v'\rangle_{x,t}(y)$, which are of great significance in turbulence modeling.  
Here, the definitions of $\langle \; \rangle_{x,t}$ and other statistic operators are given in Appendix.   

\begin{figure}
    \begin{center}
        \begin{tabular}{cc}
             \includegraphics[width=2.0in]{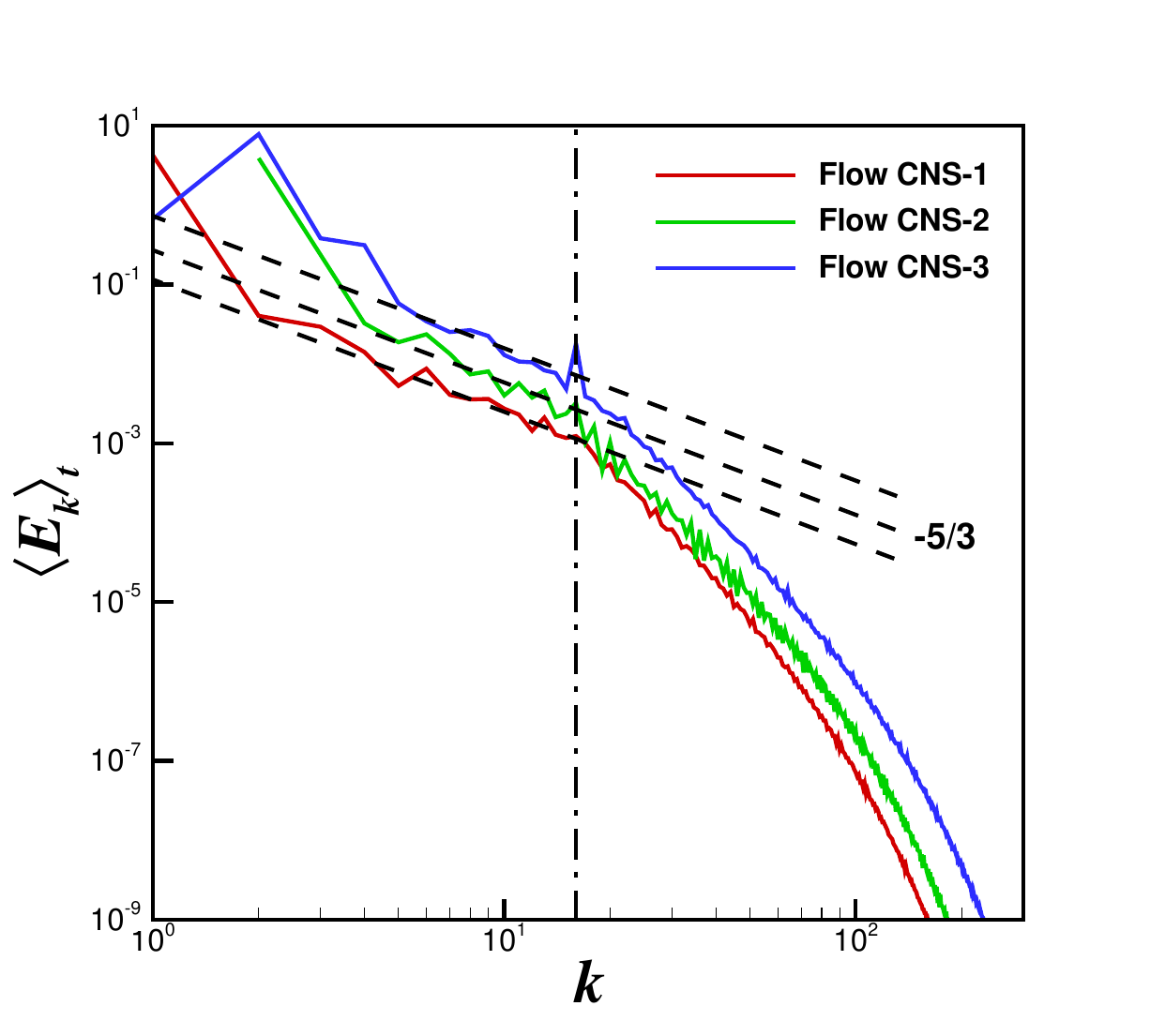}
        \end{tabular}
    \caption{Comparison of the temporal averaged kinetic energy spectra $\langle E_k \rangle_t$ of the 2D turbulent Kolmogorov flow governed by (\ref{eq_psi}) and (\ref{boundary_condition}) in the case of $n_K=16$ and $Re=2000$, given by Flow~CNS-1 (red line), Flow~CNS-2 (green line), Flow~CNS-3 (blue line), respectively, where the black dashed line corresponds to $-5/3$ power law and the black dash-dot line denotes $k=n_K=16$.}     \label{Ek_k}
    \end{center}
\end{figure} 

The above-mentioned deviations can be confirmed once again by comparing the temporal averaged kinetic energy spectra $\langle E_k \rangle_t$ of Flow~CNS-1, Flow~CNS-2, Flow~CNS-3, as shown in figure~\ref{Ek_k}: the three turbulent flows have different kinetic energy spectra, say, different distributions of kinetic energy.
It is very interesting that Flow~CNS-3 contains more kinetic energy than Flow~CNS-2, and Flow~CNS-2 contains more kinetic energy than Flow~CNS-1, although all of the three turbulent flows satisfy the -5/3 power law of Kolmogorov.   

Here we must emphasize  that the deviations between the three initial conditions considered in this investigation are rather tiny, as shown by (\ref{small-difference-IC}), which are several orders of magnitude smaller than natural/artificial stochastic disturbances. However, these tiny deviations can lead to macroscopic, obvious differences in the spatial symmetries and statistic results of the corresponding turbulent flows given by CNS, whose numerical noises are negligible compared with their true solutions throughout a long enough interval of time $t\in[0,300]$.  

All of these results provide us rigorous evidence that the corresponding Navier-Stokes turbulences of the 2D Kolmogorov flows are ultra-chaotic, say, their small disturbances {\em cannot} be neglected even from the viewpoint of {\em statistics}.         

\section{Concluding remarks and discussions}

In this paper, using the 2D turbulent Kolmogorov flow as an example, we illustrate that the Navier-Stokes turbulence (i.e. a turbulent flow governed by Navier-Stokes equations subject to initial/boundary conditions) is ultra-chaotic, say, {\em not only} its spatiotemporal trajectory {\em but also} its statistical results have sensitivity dependence on initial condition. Obviously, ultra-chaos has higher disorder than normal-chaos. This also reveals one of the reasons why turbulence is so difficult to understand.  

Before the following discussions, we should emphasize here that the ``real turbulence'' in practice is one thing, the ``Navier-Stokes turbulence'' given by Navier-Stokes equations (i.e. a mathematical model of turbulence) is a completely {\bf different} thing.   More importantly,  due to the butterfly-effect of chaos,  the ``exact'' solution of the Navier-Stokes turbulence  is one thing,  unfortunately  its   ``numerical simulation'' might be a completely {\bf different} thing.   It seems that, as long as we human being investigate turbulent flows by numerical methods, we {\bf artificially}  add a  {\bf non-negligible} influence on them.  Whether or not a mathematical model of turbulence can well describe a real turbulence needs a lots of mathematical, numerical and experimental evidences.     
   
The discovery that the Navier-Stokes turbulence might be ultra-chaotic has important meanings in theory.   From mathematical viewpoint, the situation of Navier-Stokes turbulence at an arbitrary time $t=\tau>0$ can be regarded  as a new starting point. Thus, for the ultra-chaotic Navier-Stokes turbulence, its statistic results should be sensitive to small disturbances at {\em all} time $t \geq 0$, therefore all of the natural and/or artificial, small stochastic noises should be considered.  But unfortunately, the model of Navier-Stokes turbulence indeed  neglects all stochastic disturbances for $t>0$. This certainly leads to a serious paradox in logic.   From physical viewpoint,  tiny stochastic (and unsmooth) natural/artificial noises are unavoidable in practice. Therefore, it is unreasonable to neglect the influences of small stochastic (and unsmooth) disturbances on the Navier-Stokes turbulence.  

What fundamental characteristics should a reasonable model of turbulence in general have?  This is an important  question that we had to answer, since the Navier-Stokes turbulence as a mathematical model is now even among one of the seven Millennium Prize Problem \cite{MillenniumProblem} which has attracted the attention of many mathematicians.   According to the ultra-chaotic property of the Navier-Stokes turbulence illustrated in this paper, we suggest that a reasonable turbulence model in general should have the following three fundamental characteristics:
\begin{enumerate}
\item[(A)]  physical laws of conservation should be satisfied; 

\item[(B)]  small stochastic natural/artificial disturbances should be considered;

\item[(C)]  solution should be non-differentiable. 
\end{enumerate}

Note that the so-called Landau-Lifshitz-Navier-Stokes (LLNS) equations \cite{LLNS1959}, which include the influences of stochastic thermal fluctuation, have these three fundamental characteristics. The stochastic differential  equations (such as LLNS equations) are fundamentally different from the Navier-Stokes equations: solution of the former one is continue but unsmooth, and solution of the latter one is smooth and differentiable. So, the LLNS equations might be a more reasonable model of turbulence in physics than the Navier-Stokes equations.

Recently, a multiscale approach, wave-particle turbulence simulation (WPTS) method \cite{yang2025wave1, yang2025wave2}, is developed for the numerical simulation of turbulent flows.  A  uniform  framework  of  laminar and turbulence flow can be achieved through wave-particle decomposition and the coupled evolution, where the wave stands for the cell-resolved large-scale flow structure, while the unresolved turbulent dynamics under the employed coarse grid is modeled by the stochastic particles, carrying the evolution of turbulent kinetic energy during their movement. Of course, this approach needs and is worth further investigation. Nonetheless, it satisfies the forgoing three fundamental characteristics without doubt.

Different from turbulence, laminar flows are not chaotic, i.e. spatiotemporal trajectories of laminar flows are not sensitive to small stochastic disturbances,  so that it can be well described by differentiable functions.  Thus, the Navier-Stokes equations as a mathematic model can well predict laminar flows,  and can neglect all small physical/artificial disturbances.       

Certainly, more mathematical, physical and experimental investigations should be done in the future to give more evidences to support and/or modify our above-mentioned viewpoints.

\begin{center}
{\bf\large Appendix\\ Some definitions and measures}    \label{Key_measures}
\end{center}

For the sake of simplicity, the definitions of some statistic operators are briefly described below.
The spatial average is defined by
\begin{align}
& \langle\,\,\rangle_A=\frac{1}{4\pi^2}\int^{2\pi}_0\int^{2\pi}_0 dxdy,       \label{average_A}
\end{align}
the temporal average is defined by
\begin{align}
& \langle\,\,\rangle_{t}=\frac{1}{T_2-T_1}\int^{T_2}_{T_1} dt,       \label{average_t}
\end{align}
the spatiotemporal average (along the $x$ direction) is defined by
\begin{align}
& \langle\, \,\rangle_{x,t}=\frac{1}{2\pi (T_2-T_1)}\int^{2\pi}_0\int^{T_2}_{T_1} \, \mathrm{d}x \mathrm{d}t,       \label{average_xt}
\end{align}
and the overall spatiotemporal average is defined by
\begin{align}
& \langle\,\,\rangle=\frac{1}{4\pi^2 (T_2-T_1)}\int^{2\pi}_0\int^{2\pi}_0\int^{T_2}_{T_1} dxdydt.       \label{average_all}
\end{align}
For an interval of time corresponding to the relatively stable state of turbulence, $T_1=100$ and $T_2=300$ are chosen in this paper.

%For the turbulent two-dimensional Kolmogorov flow considered in this paper, vorticity is given by the stream function
%\begin{align}
%& \omega(x,y,t)=\nabla^{2}\psi(x,y,t).       \label{vorticity}
%\end{align}
%We also focus on the kinetic energy
%\begin{align}
%& E(x,y,t) = \frac{1}{2}[u^2(x,y,t)+v^2(x,y,t)],    \label{kinetic_energy}
%\end{align}
%enstrophy
%\begin{align}
%& \Omega(x,y,t) = \frac{1}{2}\,\omega^2(x,y,t),    \label{enstrophy}
%\end{align}
%the kinetic energy dissipation rate
%\begin{align}
%& D(x,y,t)=\frac{1}{2Re}\sum_{i,j=1,2}\big [ \partial_iu_j(x,y,t)+\partial_ju_i(x,y,t) \big ]^2,    \label{dissipation_rate}
%\end{align}
%and enstrophy dissipation rate
%\begin{align}
%& D_\Omega(x,y,t)=\frac{1}{Re}|\nabla\omega(x,y,t)|^2,    \label{enstrophy_dissipation_rate}
%\end{align}
%where $u_1(x,y,t)=u(x,y,t)$, $u_2(x,y,t)=v(x,y,t)$, $\partial_1=\partial /\partial x$, and $\partial_2=\partial /\partial y$.

The stream function can be expanded as the Fourier series
\begin{align}
& \psi(x,y,t)\approx\sum^{\lfloor N/3 \rfloor}_{\,m=-\lfloor N/3 \rfloor}\sum^{\lfloor N/3 \rfloor}_{\,n=-\lfloor N/3 \rfloor}\Psi_{m,n}(t) \exp(\mathbf{i}\,mx)\exp(\mathbf{i}\,ny),       \label{Fourier}
\end{align}
where $m$, $n$ are integers, $\lfloor\,\,\rfloor$ stands for a floor function, $\mathbf{i}=\sqrt{-1}$ denotes the imaginary unit, and for dealiasing $\Psi_{m,n}=0$ is imposed for wavenumbers outside the above domain $\sum$. Note that for the real number $\psi$, $\Psi_{-m,-n}=\Psi^*_{m,n}$ must be satisfied, where $\Psi^*_{m,n}$ is the conjugate of $\Psi_{m,n}$.
Therefore, the kinetic energy spectrum is defined as
\begin{align}
& E_k(t)=\sum_{k-1/2 \leq \sqrt{m^2+n^2} < k+1/2}\frac{1}{2}\,(m^2+n^2)\mid \Psi_{m,n}(t) \mid^2,       \label{kinetic_energy_spectrum}
\end{align}
where the wave number $k$ is a non-negative integer.

By Reynolds decomposition, the velocity $\mathbf{u}=(u, v)$ is divided into mean and fluctuating quantities
\begin{align}
& \mathbf{u}=\langle\, \mathbf{u} \,\rangle_{x,t}+\mathbf{u}',       \label{decomposition}
\end{align}
considering the homogeneity in $x$ direction because of the form of external force, which also leads to $\langle\, v \,\rangle_{x,t}=0$.

\vspace{0.5cm}

\section*{Acknowledgements}
The calculations were performed on ``Tianhe New Generation Supercomputer'', National Supercomputer Center in Tianjing, China. This work is partly supported by National Natural Science Foundation of China (No. 12302288 \& 91752104),  State Key Laboratory of Ocean Engineering, and Dept. of Mathematics, Hong Kong University of Science and Technology.

%%Vancouver style references.
\bibliographystyle{elsarticle-num}
\bibliography{Kolmogorov}

\end{document}